\RequirePackage{fix-cm} 
\documentclass[a4paper, twoside, reqno, dvips, 12pt]{amsart}
\usepackage{fixltx2e}   

\usepackage{etex}

\usepackage[latin1]{inputenc}
\usepackage[T1]{fontenc}

\usepackage{eucal}
\usepackage{esint}
\usepackage{dsfont}
\usepackage{xspace}
\usepackage{amsgen}
\usepackage{amsthm}
\usepackage{amssymb}
\usepackage{amsmath}
\usepackage{upgreek}
\usepackage{wasysym}
\usepackage{amsfonts}
\usepackage{stmaryrd}
\usepackage{mathtools}

\usepackage{mathrsfs}
\DeclareMathAlphabet{\mathscrbf}{OMS}{mdugm}{b}{n}

\usepackage{a4wide}

\usepackage[dvipsnames, table]{xcolor}
\definecolor{bckg}{RGB}{20.8, 20.8, 20.8}
\definecolor{oneblue}{rgb}{0.0, 0.0, 0.85}
\definecolor{Lightblue}{RGB}{214, 214, 214}
\definecolor{bluepigment}{rgb}{0.2, 0.2, 0.6}
\definecolor{charcoal}{rgb}{0.21, 0.27, 0.31}
\definecolor{denimblue}{rgb}{0.08, 0.38, 0.74}
\definecolor{Lightgray}{rgb}{0.89, 0.89, 0.89}
\definecolor{darkgrey}{rgb}{0.273, 0.281, 0.30}
\definecolor{darkelectricblue}{rgb}{0.33, 0.41, 0.47}

\usepackage{psfrag}
\usepackage{graphicx}
\usepackage{subfigure}
\usepackage{morefloats}
\usepackage{indentfirst}

\usepackage{acronym}
\usepackage{microtype}
\usepackage[labelsep=period,%
            labelfont={bf,sf,color=bluepigment},%
            justification=raggedright]{caption}

\usepackage[perpage, symbol]{footmisc}

\usepackage[usenames, dvipsnames]{pstricks}
\usepackage{epsfig}
\usepackage{pst-grad} 
\usepackage{pst-plot} 

\usepackage[colorlinks,
           urlcolor=oneblue,
           linkcolor=denimblue,
           citecolor=NavyBlue,
           bookmarksopen=false,
           pdfpagemode=UseNone,
           pagebackref]{hyperref}

\usepackage[sort&compress, comma, square, numbers]{natbib}

\usepackage[explicit]{titlesec}

\titleformat{\section}
  {\color{NavyBlue}\Large\sffamily\bfseries}
  {}
  {0em}
  {\colorbox{bckg!5}{\parbox{\dimexpr\linewidth-2\fboxsep\relax}{\centering\thesection. #1}}}
  [\vspace*{0.33em}]

\titleformat{name=\section,numberless}
  {\color{NavyBlue}\Large\sffamily\bfseries}
  {}
  {0.0em}
  {\colorbox{bckg!10}{\parbox{\dimexpr\linewidth-2\fboxsep\relax}{\centering#1}}}
  [\vspace*{0.33em}]

\titleformat{\subsection}
  {\color{NavyBlue}\large\sffamily\bfseries}
  {}
  {0.0em}
  {\colorbox{bckg!5}{\parbox{\dimexpr\linewidth-2\fboxsep\relax}{\centering\thesubsection. #1}}}
  [\vspace*{0.33em}]

\titleformat{name=\subsection,numberless}
  {\color{NavyBlue}\Large\sffamily\bfseries}
  {}
  {0em}
  {\colorbox{bckg!10}{\parbox{\dimexpr\linewidth-2\fboxsep\relax}{\centering#1}}}
  [\vspace*{0.33em}]

\titleformat{\subsubsection}
  {\color{bluepigment}\sffamily\normalsize\bfseries}
  {\thesubsubsection}
  {0.5em}
  {#1}
  [\vspace*{0.33em}]

\titleformat{\paragraph}[runin]
  {\color{bluepigment}\sffamily\small\bfseries}
  {}
  {0em}
  {#1}

\titlespacing{\section}{1.0em}{1.5em plus 2pt minus 2pt}%
{1.0em plus 2pt minus 2pt}[0em]
\titlespacing{\subsection}{1.0em}{1.5em plus 2pt minus 2pt}%
{1.0em}[0em]
\titlespacing{\subsubsection}{1.0em}{1.5em plus 2pt minus 2pt}%
{1.0em plus 2pt minus 2pt}[0em]

\usepackage{titletoc}

\contentsmargin{0.5em}
\newlength{\tocsep} 
\titlecontents{section}[\tocsep]
  {\addvspace{10pt}\bfseries\sffamily}
  {\contentslabel[\thecontentslabel]{\tocsep}}
  {}
  {\ \titlerule*[0.75pc]{.}\ \thecontentspage}
  []
\titlecontents{subsection}[\tocsep]
  {\addvspace{8pt}\sffamily}
  {\contentslabel[\thecontentslabel]{\tocsep}}
  {}
  {\ \titlerule*[0.5pc]{.}\ \thecontentspage}
  []
\titlecontents*{subsubsection}[\tocsep]
  {\addvspace{2pt}\footnotesize\sffamily}
  {}
  {}
  {\ \titlerule*[0.35pc]{.}\ \thecontentspage}
  [\\*]

\makeatletter
\def\@setauthors{%
  \begingroup
  \def\thanks{\protect\thanks@warning}%
  \trivlist
  \centering\footnotesize \@topsep30\p@\relax
  \advance\@topsep by -\baselineskip
  \item\relax
  \author@andify\authors
  \def\\{\protect\linebreak}%
  \textsc{\normalsize\textcolor{darkelectricblue}{\authors}}%
  \ifx\@empty\contribs
  \else
    ,\penalty-3 \space \@setcontribs
    \@closetoccontribs
  \fi
  \endtrivlist
  \endgroup
}
\def\@settitle{\begin{center}%
  \baselineskip14\p@\relax
    \bfseries
    \textsc{\Large\textcolor{charcoal}{\@title}}
  \end{center}%
}
\makeatother

\usepackage{enumitem}
\setlist[description]{%
  topsep=30pt,               
  itemsep=5pt,               
  font={\bfseries\sffamily\color{NavyBlue}}, 
}

\usepackage{fancyhdr}
\usepackage{lastpage}

\newcommand*\Title{\textcolor{bluepigment}{On weakly singular travelling CG waves}}
\newcommand*\Authors{\textcolor{bluepigment}{D.~Mitsotakis, D.~Dutykh \etal}}
\newcommand*{\plogo}{\textcolor{gray}{{\texttt{arXiv.org} / \textsc{hal}}}} 

\numberwithin{equation}{section}

\newcommand{\E}{\mathcal{E}}
\newcommand{\ud}{\mathrm{d}}
\newcommand{\ue}{\mathrm{e}}
\newcommand{\Ke}{\mathcal{K}}
\newcommand{\Rr}{\mathscr{R}}
\newcommand{\Hh}{\mathscr{H}}
\newcommand{\Qq}{\mathscr{Q}}
\newcommand{\Fr}{\mathsf{Fr}}
\newcommand{\Bo}{\mathsf{Bo}}

\renewcommand{\L}{\mathscr{S}}
\newcommand{\Vc}{\mathcal{V}_{\,c}}
\newcommand{\Vg}{\mathcal{V}_{\,g}}
\newcommand{\const}{\mathrm{const}}

\renewcommand{\sim}{\thicksim}

\newcommand{\abs}[1]{\lvert\, #1\, \rvert}

\newcommand{\pd}[2]{\frac{\partial\/ #1}{\partial\/ #2}}

\newcommand{\eqdef}{\mathop{\stackrel{\,\mathrm{def}}{:=}\,}}

\newcommand{\ie}{\emph{i.e.}\xspace}

\newcommand{\etal}{\emph{etal.}\xspace}

\newcommand{\half}{{\textstyle{1\over2}}}
\newcommand{\third}{{\textstyle{1\over3}}}

\begin{document}

\title[\Title]{On weakly singular and fully nonlinear travelling shallow capillary--gravity waves in the critical regime}

\author[D.~Mitsotakis]{Dimitrios Mitsotakis}
\address{\textbf{D.~Mitsotakis:} Victoria University of Wellington, School of Mathematics, Statistics and Operations Research, PO Box 600, Wellington 6140, New Zealand}
\email{dmitsot@gmail.com}
\urladdr{http://dmitsot.googlepages.com/}

\author[D.~Dutykh]{Denys Dutykh$^*$}
\address{\textbf{D.~Dutykh:} LAMA, UMR 5127 CNRS, Universit\'e Savoie Mont Blanc, Campus Scientifique, F-73376 Le Bourget-du-Lac Cedex, France}
\email{Denys.Dutykh@univ-savoie.fr}
\urladdr{http://www.denys-dutykh.com/}
\thanks{$^*$ Corresponding author}

\author[A.~Assylbekuly]{Aydar Assylbekuly}
\address{\textbf{A.~Assylbekuly:} Khoja Akhmet Yassawi International Kazakh--Turkish University, Faculty of Natural Science, Department of Mathematics, 161200 Turkestan, Kazakhstan}
\email{asylbekuly@mail.ru}

\author[D.~Zhakebayev]{Dauren Zhakebayev}
\address{\textbf{D.~Zhakebayev:} Al-Farabi Kazakh National University, Faculty of Mechanics and Mathematics, Department of Mathematical and Computer Modelling, 050000 Almaty, Kazakhstan}
\email{daurjaz@mail.ru}

\keywords{capillary--gravity waves; peakons; nonlinear dispersive waves}


\begin{titlepage}
\thispagestyle{empty} 
\noindent
{\Large Dimitrios \textsc{Mitsotakis}}\\
{\it\textcolor{gray}{Victoria University of Wellington, New Zealand}}
\\[0.02\textheight]
{\Large Denys \textsc{Dutykh}}\\
{\it\textcolor{gray}{CNRS, Universit\'e Savoie Mont Blanc, France}}
\\[0.02\textheight]
{\Large Aydar \textsc{Assylbekuly}}\\
{\it\textcolor{gray}{Khoja Akhmet Yassawi International Kazakh--Turkish University, Kazakhstan}}
\\[0.02\textheight]
{\Large Dauren \textsc{Zhakebayev}}\\
{\it\textcolor{gray}{Al-Farabi Kazakh National University, Kazakhstan}}
\\[0.16\textheight]

\colorbox{Lightblue}{
  \parbox[t]{1.0\textwidth}{
    \centering\huge\sc
    \vspace*{0.7cm}
    
    \textcolor{bluepigment}{On weakly singular and fully nonlinear travelling shallow capillary$-$gravity waves in the critical regime}

    \vspace*{0.7cm}
  }
}

\vfill 

\raggedleft     
{\large \plogo} 
\end{titlepage}


\newpage
\thispagestyle{empty} 
\par\vspace*{\fill}   
\begin{flushright} 
{\textcolor{denimblue}{\textsc{Last modified:}} \today}
\end{flushright}


\newpage
\maketitle
\thispagestyle{empty}


\begin{abstract}

In this Letter we consider long capillary--gravity waves described by a fully nonlinear weakly dispersive model. First, using the phase space analysis methods we describe all possible types of localized travelling waves. Then, we especially focus on the critical regime, where the surface tension is exactly balanced by the gravity force. We show that our long wave model with a critical Bond number admits stable travelling wave solutions with a singular crest. These solutions are usually referred to in the literature as \emph{peakons} or \emph{peaked solitary waves}. They satisfy the usual speed-amplitude relation, which coincides with Scott-Russel's empirical formula for solitary waves, while their decay rate is the same regardless their amplitude. Moreover, they can be of depression or elevation type independent of their speed. The dynamics of these solutions are studied as well. \\

\bigskip
\noindent \textbf{\keywordsname:} capillary--gravity waves; peakons; nonlinear dispersive waves \\

\bigskip
\noindent \textbf{MSC:} \subjclass[2010]{ 76B25 (primary), 76B15, 35Q51, 35C08 (secondary)}\smallskip \\
\noindent \textbf{PACS:} \subjclass[2010]{ 47.35.Bb (primary), 47.35.Pq, 47.35.Fg (secondary)}

\end{abstract}


\newpage
\tableofcontents
\thispagestyle{empty}


\clearpage
\section{Introduction}

In this Letter we investigate further the problem of hydrodynamic wave propagation over a horizontal impermeable bottom while we focus on a very particular regime of long capillary--gravity waves.  Consider a two-dimensional \textsc{Cartesian} coordinate system $Oxy$ where its horizontal axis coincides with the still water level $y\ =\ 0$. A layer of a perfect incompressible fluid is bounded from below by a flat impermeable bottom $y\ =\ -d$ and from above by the free surface $y\ =\ \eta\,(x,\,t)$. The fluid density is assumed to be constant $\rho\ >\ 0$. The total water depth is denoted by $h\,(x,\,t) \eqdef d\ +\ \eta\,(x,\,t)$. Since the bottom is flat (\ie $d\ =\ \const$), we can equivalently replace the derivatives of the free surface elevation by the same derivatives of the total water depth, \ie $\eta_{\,t}\ \equiv\ h_{\,t}$, $\eta_{\,x}\ \equiv\ h_{\,x}$. We shall use this property below.

In this derivation we follow the main lines of our previous work \cite{Clamond2015c}. According to the \textsc{Young}--\textsc{Laplace} law, the pressure $p$ jump across the interface is given by the following relation:
\begin{equation*}
  \llbracket\, p\, \rrbracket = -\sigma\,\biggl[\,\frac{\eta_{\,x}}{\sqrt{1\ +\ \eta_{\,x}^{\,2}}}\,\biggr]_{\,x} = -\sigma\;\frac{\eta_{\,x\,x}}{\bigl(1\ +\ \eta_{\,x}^{\,2}\bigr)^{\,3/2}}\,,
\end{equation*}
where $\sigma$ represents the surface tension. In this study we apply the small (free surface's) slope approximation to obtain $\llbracket\, p\, \rrbracket \approx -\sigma\,\eta_{\,x\,x}$. This pressure jump appears in the water wave problem through the \textsc{Cauchy}--\textsc{Lagrange} integral which serves as the dynamic boundary condition. Below we shall return to the surface tension effects by considering their potential energy since it allows to achieve easier our goals.

In order to derive model equations for gravity-capillary surface water waves we have to choose an ansatz to flow's structure and compute the system energy. Usually these model equations can be obtained if the horizontal velocity $u(x,\,y,\,t)$ is approximated by the depth-averaged fluid velocity $\bar{u}(x,\, t)$, and the vertical velocity is chosen to satisfy identically the incompressibility and bottom impermeability:
\begin{equation*}
  u(x,\,y,\,t)\ \approx\ \bar{u}(x,\,t)\,, \quad
  v(x,\,y,\,t)\ \approx\ -(y\ +\ d)\,\bar{u}_{\,x}(x,\,t)\,.
\end{equation*}
Below we shall omit over bars in the notation since we work only with the depth-averaged velocity.

The various forms of energies for the specific fluid flow are estimated bellow: The kinetic energy $\Ke$ consists of the hydrostatic and non-hydrostatic corrections:
\begin{equation*}
  \Ke = \int_{t_1}^{t_2}\int_{x_1}^{x_2} \rho\,\biggl[\,\frac{h\,u^{\,2}}{2}\ +\ \frac{h^{\,3}\,u_{\,x}^{\,2}}{6}\,\biggr]\;\ud x\;\ud t\,.
\end{equation*}
The potential energy consists of the gravity
\begin{equation*}
  \Vg = \int_{t_1}^{t_2}\int_{x_1}^{x_2}\frac{\rho\,g\,h^{\,2}}{2}\;\ud x\;\ud t\,,
\end{equation*}
and capillary contributions:
\begin{equation*}
  \Vc = \int_{t_1}^{t_2}\int_{x_1}^{x_2}\rho\,\tau\;\Bigl[\,\sqrt{1 + h_{\,x}^{\,2}} - 1\,\Bigr]\;\ud x\;\ud t\ \approx\ \frac{1}{2}\;\int_{t_1}^{t_2}\int_{x_1}^{x_2}\rho\,\tau\;h_{\,x}^{\,2}\;\ud x\;\ud t\,,
\end{equation*}
to which we applied the small slope approximation (and we introduced another physical constant $\tau \eqdef \frac{\sigma}{\rho}$). For more details on the derivation of energies $\Ke$ and $\Vg$ we refer to \cite{Clamond2015c}. Now we can assemble the action integral:
\begin{equation*}
  \L \eqdef \Ke - \Vg - \Vc + \int_{t_1}^{t_2}\int_{x_1}^{x_2} \rho\,\bigl[\,h_{\,t}\ +\ [h\,u]_{\,x}\,\bigr]\,\phi\;\ud x\;\ud t\,,
\end{equation*}
where we enforced the mass conservation by introducing a \textsc{Lagrange} multiplier $\phi(x,\,t)\,$. By applying the \textsc{Hamilton}--\textsc{Ostrogradsky} variational principle and eliminating the \textsc{Lagrange} multiplier $\phi(x,\,t)$ from the equations, we arrive at the following system of equations:
\begin{align}\label{eq:serre1}
  h_{\,t} + [h\,u]_{\,x} &= 0\,, \\
  u_{\,t} + u\,u_{\,x} + g\,h_{\,x} &= \frac{1}{3\,h}\;\bigl[\,h^{3}(u_{\,x\,t} + u\,u_{\,x\,x} - u_{\,x}^{\,2})\,\bigr]_{\,x}\ +\ \tau\, h_{\,x\,x\,x}.\label{eq:serre2}
\end{align}
These are the celebrated \textsc{Serre}--\textsc{Green}--\textsc{Naghdi} (SGN) equations with weak\footnote{The word `weak' comes from the fact that we applied small slope approximation.} surface tension effects, \cite{Dias2010, Lannes2013}. The full list of (physical) conservation laws is given below. The mass conservation was already given in equation \eqref{eq:serre1}.  The remaining identities are given below:
\begin{align}\label{eq:cons1}
  \biggl[\,u - \frac{(h^{\,3}\,u_{\,x})_{\,x}}{3\,h}\,\biggr]_{\,t}\ + \ \biggl[\,\frac{u^{\,2}}{2} + g\,h - \frac{h^{\,2}\,u_{\,x}^{\,2}}{2} - \frac{u\,(h^{\,3}\,u_{\,x})_{\,x}}{3\,h} - \tau\,h_{\,x\,x}\,\biggr]_{\,x} &= 0\,, \\
  [h\,u]_{\,t} + \bigl[\,h\,u^{\,2} + \half\,g\,h^{\,2} + \third\,h^{\,2}\,\gamma - \tau\,\Rr\,\bigr]_{\,x} &= 0\,, \label{eq:cons3}
\end{align}
where we introduced for the sake of notation compactness two quantities:
\begin{align*}
  \gamma &\eqdef h\,\bigl[\,u_{\,x}^{\,2}\ -\ u_{\,x\,t}\ -\ u\,u_{\,x\,x}\,\bigr]\,, \\
  \Rr &\eqdef h\,h_{\,x\,x}\ -\ \half\,h_{\,x}^{\,2}\,.
\end{align*}
The quantity $\gamma$ has a physical sense of the vertical acceleration of fluid particles computed at the free surface. Some of the conservation laws shall be used below to study travelling waves to the SGN system \eqref{eq:serre1}-\eqref{eq:serre2}.

The conservation of energy can be written in the form:
\begin{equation}\label{eq:energy}
\Hh_t+\Qq_x \ = \ 0\,,
\end{equation}
where
\begin{align*}
  \Hh\ &\eqdef\ \frac{hu^{2}}{2}\ +\ \frac{h^{3}u_{x}^{2}}{6}\ +\ \frac{gh^{2}}{2}\ +\ \frac{\tau}{2}h_{x}^{2}\,, \\
  \Qq\ &\eqdef\ \Bigl(\frac{u^{2}}{2}\ +\ \frac{h^{2}u_{x}^{2}}{6}\ +\ gh\ +\ \frac{h\gamma}{3}\ -\ \tau h_{xx}\Bigr)hu\ +\ \tau h_{x}(hu)_{x}\,,
\end{align*}
with $\Hh$ the approximation of the total energy and $\Qq$ the energy flux.

The nature of the solutions of the SGN system depend on  the  parameter $\tau$ and they have been studied for the values of $\tau\ \neq\ 1/3\,$. It is know that for $\tau\ <\ 1/3$ the system admits classical solitary waves of elevation while for $\tau\ >\ 1/3$ there are only solitary waves of depression. In the rest of the paper we study in detail the traveling wave solutions of the SGN equation with special emphasis in the critical case $\tau\ =\ 1/3\,$.


\section{Travelling wave solutions}

In this Section we focus on a special class of solutions --- the so-called travelling waves. The main simplifying circumstance is that the flow becomes steady in the frame of reference moving with the wave. Thus, it allows to analyze Ordinary Differential Equations (ODEs) instead of working with Partial Differential Equations (PDEs). We substitute the following solution \emph{ansatz} into all equations:
\begin{equation*}
  u\,(x,\,t)\ =\ u(\xi)\,, \quad h\,(x,\,t)\ =\ h(\xi)\,, \quad
  \xi\ \eqdef\ x\ -\ c\,t\,,
\end{equation*}
where $c\ >\ 0$ is the wave speed\footnote{In other words, we consider waves moving in the rightward direction, without loosing generality.}. Moreover, we focus on localized solutions of this type --- the so-called \emph{solitary waves}. They satisfy the following boundary conditions:
\begin{equation*}
  h^{(n)}\,(\xi) \to 0\,, \qquad u^{(n)}\,(\xi) \to 0\,, \quad \mbox{ as } \quad \xi \to \infty\,,
\end{equation*}
$n\ =\ 1,\,2,\,\ldots\,$. For the total depth and horizontal velocity profiles (\ie $n\ =\ 0$) we have the following boundary conditions:
\begin{equation*}
  h\,(\xi) \to d\,, \quad u\,(\xi) \to -c\,, \quad \mbox{ as } \quad \xi \to \infty\,.
\end{equation*}

The mass conservation equation \eqref{eq:serre1} readily yields a relation between $u$ and $h\,$:
\begin{equation}\label{eq:massSteady}
  u\,(\xi) = -\,\frac{c\,d}{h\,(\xi)}\,.
\end{equation}
By substituting these relations into the conservation laws \eqref{eq:cons1}, \eqref{eq:cons3} and taking a linear combination of these two equations leads to the following \emph{implicit} ODE $\E(h^{\,\prime},\,h) = 0$ for the total water depth:
\begin{multline}\label{eq:master}
  \E(h^{\,\prime},\,h) \eqdef \frac{\Fr\,(h^{\,\prime})^{\,2}}{3}\ -\ \Bo\,(h^{\,\prime})^{\,2}\;\frac{h}{d}\ -\ \Fr\\
  +\ \frac{(2\,\Fr\ +\ 1)\,h}{d}\ -\ \frac{(\Fr\ +\ 2)\,h^{\,2}}{d^{\,2}}\ +\ \frac{h^{\,3}}{d^{\,3}}\ =\ 0\,,
\end{multline}
where we introduced two dimensionless numbers:
\begin{description}
  \item[$\Fr\ \eqdef\ \frac{c^{\,2}}{g\,d}$] the \textsc{Froude} (also known as E\"otv\"os) number
  \item[$\Bo\ \eqdef\ \frac{\tau}{g\,d^{\,2}}\ \equiv\ \frac{\sigma}{\rho\,g\,d^{\,2}}$] the \textsc{Bond} number.
\end{description}
The details on the derivation of the master equation \eqref{eq:master} with the full surface tension term can be found in \cite{Clamond2016b}. Let us compute the partial derivatives of the function $\E(h^{\,\prime},\,h)$:
\begin{align*}
  \pd{\E}{h}\ &=\ -\frac{\Bo\,(h^{\,\prime})^{\,2}}{d} + \frac{2\,\Fr + 1}{d} - \frac{2\,(\Fr + 2)\,h}{d^{\,2}} + \frac{3\,h^{\,3}}{d^{\,3}}\,, \\
  \pd{\E}{h^{\,\prime}}\ &=\ \frac{2\,\Fr\,h^{\,\prime}}{3} - 2\,\Bo\,h^{\,\prime}\;\frac{h}{d}\,.
\end{align*}
In particular, one can see that the derivative $\pd{\E}{h^{\,\prime}}$ may vanish. It implies that singularities may exist \cite{Arnold1996}. Equation \eqref{eq:master} can be seen as an algebraic relation between two variables $h$ and $h^{\,\prime}$ which defines implicitly an algebraic curve in the phase plane $(h,\,h^{\,\prime})\,$. A solitary wave necessarily `lives' on this algebraic variety. Namely, it has to be a homoclinic orbit departing and returning to the point $(d,\,0)$ by respecting the orientation of branches (\ie $h$ is obviously increasing where $h^{\,\prime}\ >\ 0$ and vice versa). For the case of full surface tension all possible topologies of these curves were analyzed in \cite{Clamond2016b}. Here we repeat some parts of this analysis for the weak case. Just by looking at equation \eqref{eq:master} we can infer several properties of solitary waves to the SGN equations:
\begin{itemize}
  \item Solitary waves are symmetric with respect to the crest
  \item For solitary waves with a smooth crest the following speed-amplitude relation holds:
  \begin{equation}\label{eq:speedampl}
    \Fr\ =\ 1\ +\ \frac{a}{d}\,,
  \end{equation}
  where $a$ is the wave amplitude, \ie $h\,(0)\ =\ a\ +\ d\,$
  \item By performing \textsc{McCowan}'s analysis \cite{McCowan1891} we obtain the following `dispersion relation' for solitary waves:
  \begin{equation}\label{eq:disprel}
    \Fr\ =\ \frac{3\,\bigl(1\ -\ \Bo\,(\kappa\,d)^{\,2}\bigr)}{3\ -\ (\kappa\,d)^{\,2}}\,,
  \end{equation}
  where $\kappa$ is the solitary wave decay rate, \ie $h\,(\xi)\ \sim\ d\ +\ a\,\ue^{-\kappa\,\xi}\,$.
\end{itemize}
We would like to comment on the last point. If $\Bo\ \equiv\ 1/3$ (the critical case) then we have necessarily that $\Fr\ =\ 1$ \emph{or} $\kappa\,d\ =\ \sqrt{3}\,$. Moreover, the critical case is necessarily \emph{non-dispersive}.


\subsection{Phase plane analysis}

For fixed (meaningful) values of parameters $\Fr$ and $\Bo$ we can construct (and plot) the corresponding phase plane diagram using, for example, \texttt{algcurves} package from \textsc{Maple}, which offers tools for studying one-dimensional algebraic curves defined by multi-variate polynomials. This package detects all the branches and, thus, the results are certified. Travelling wave solutions (both solitary and periodic) to several two-component systems (including two-component \textsc{Camassa}--\textsc{Holm} and modified \textsc{Green}--\textsc{Naghdi} equations) were completely classified in \cite{Dutykh2016} using similar phase plane considerations.

\begin{figure}
  \centering
  \subfigure[]{\includegraphics[width=0.48\textwidth]{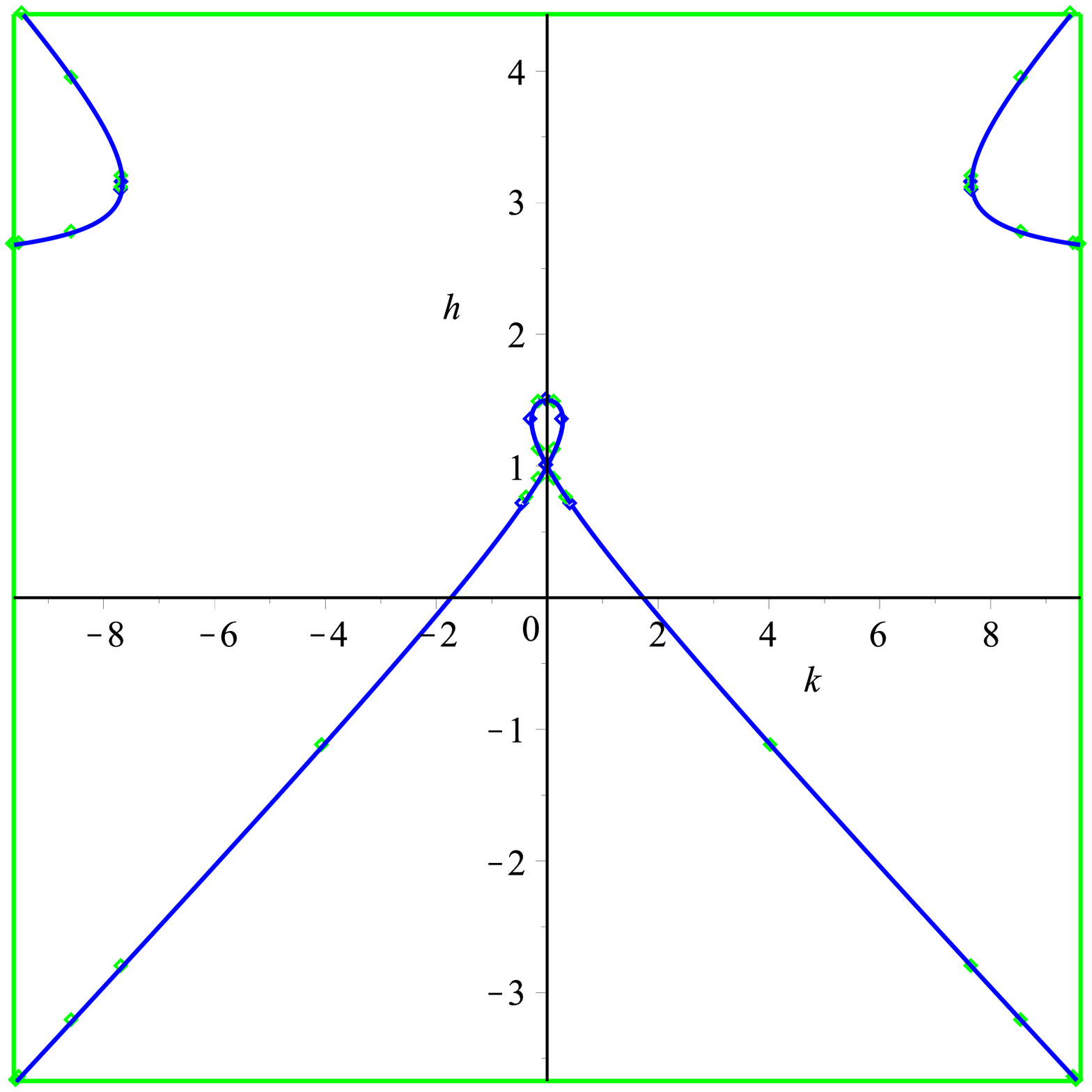}}
  \subfigure[]{\includegraphics[width=0.48\textwidth]{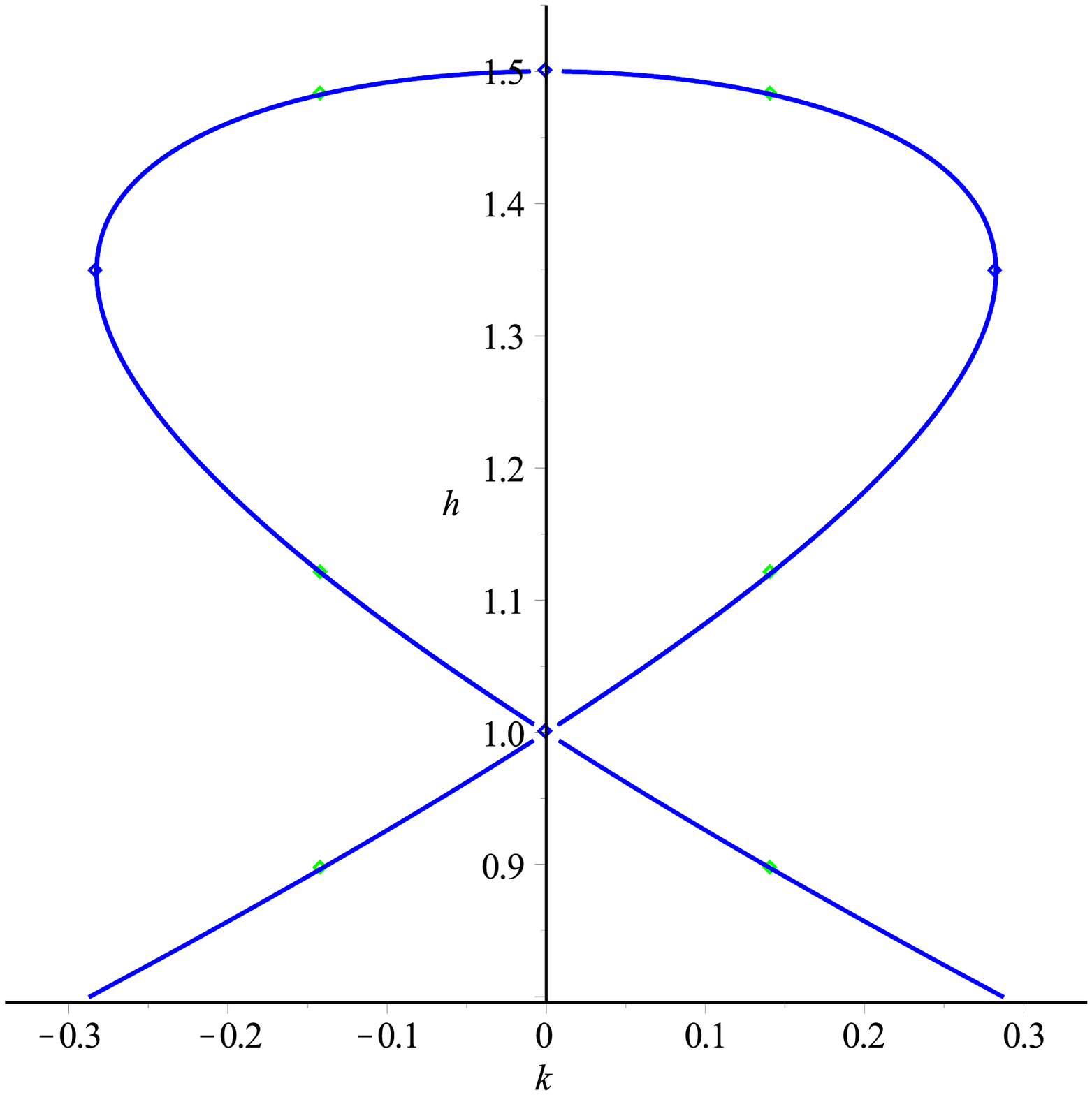}}
  \caption{\small\em Sub-critical case: the algebraic variety (a) and a zoom on the solitary wave (b) for parameters $\Fr\ =\ 1.5\ >\ 1$ and $\Bo\ =\ 0.2\ <\ \third\,$. On this and all other phase plane diagrams $k\ \equiv\ h^{\,\prime}\,$.}
  \label{fig:subcrit}
\end{figure}

\begin{figure}
  \centering
  \subfigure[]{\includegraphics[width=0.48\textwidth]{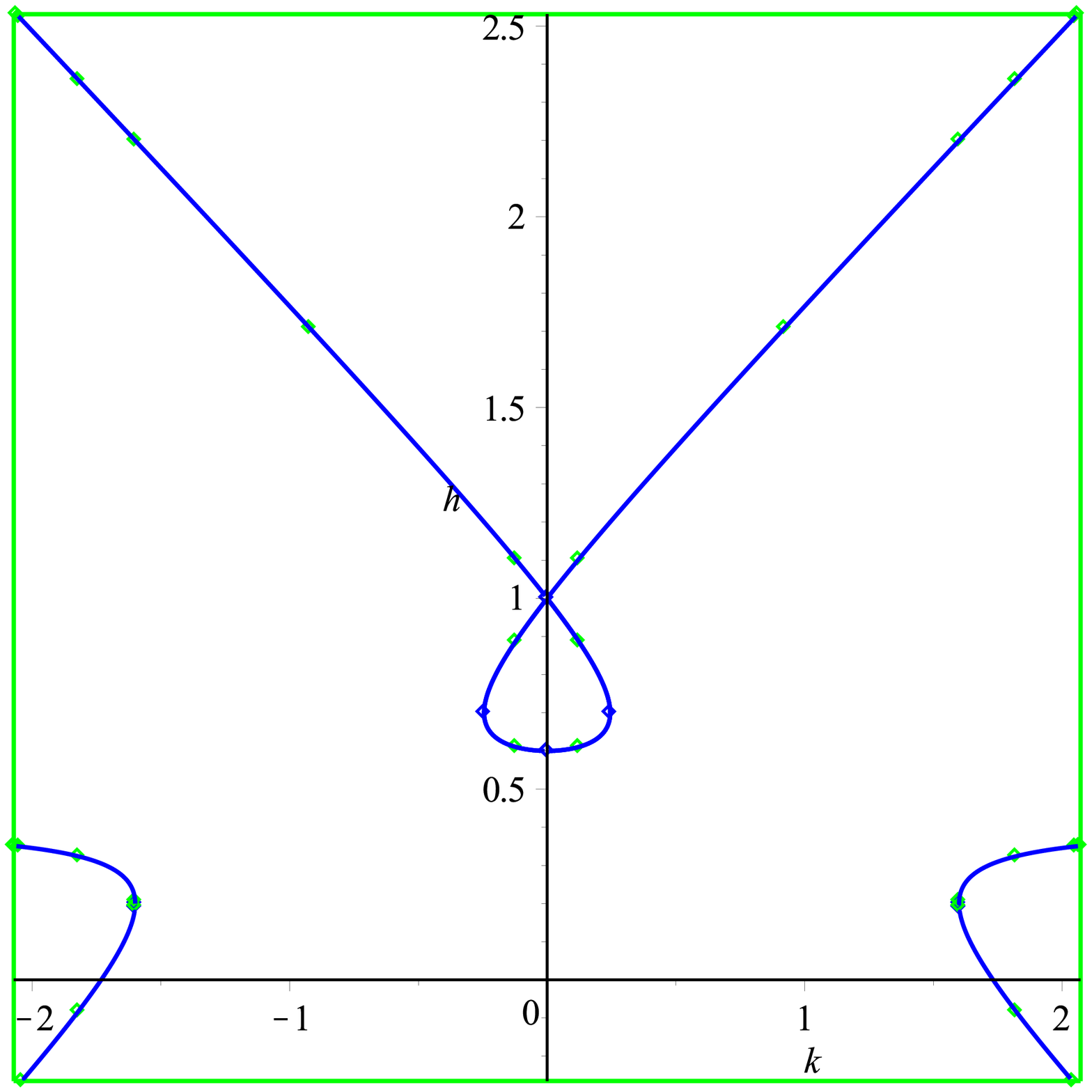}}
  \subfigure[]{\includegraphics[width=0.48\textwidth]{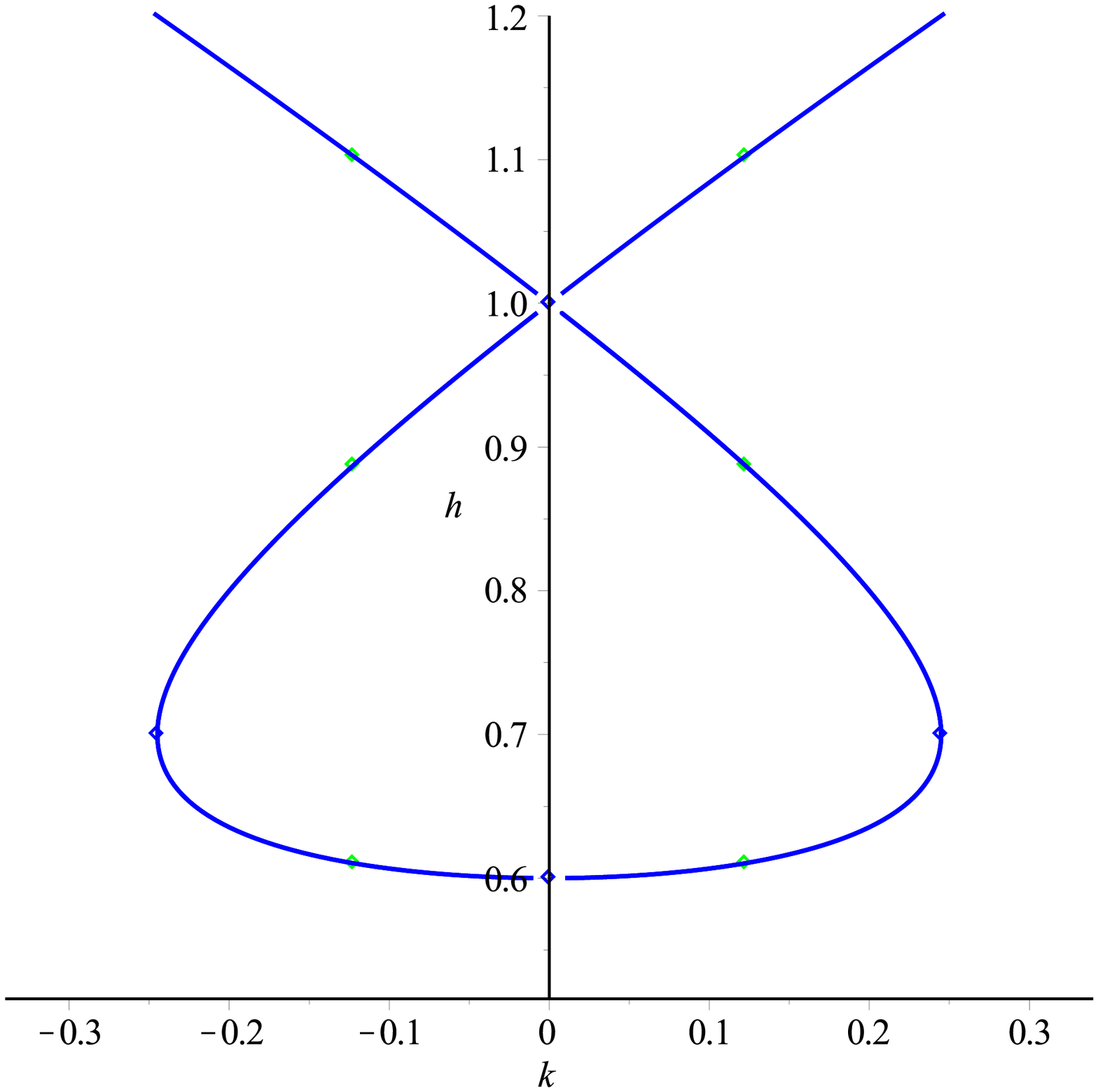}}
  \caption{\small\em Super-critical case: the algebraic variety (a) and a zoom on the solitary wave (b) for parameters $\Fr\ =\ 0.6\ <\ 1$ and $\Bo\ =\ 0.5\ >\ \third\,$.}
  \label{fig:supercrit}
\end{figure}

\begin{figure}
  \centering
  \subfigure[]{\includegraphics[width=0.48\textwidth]{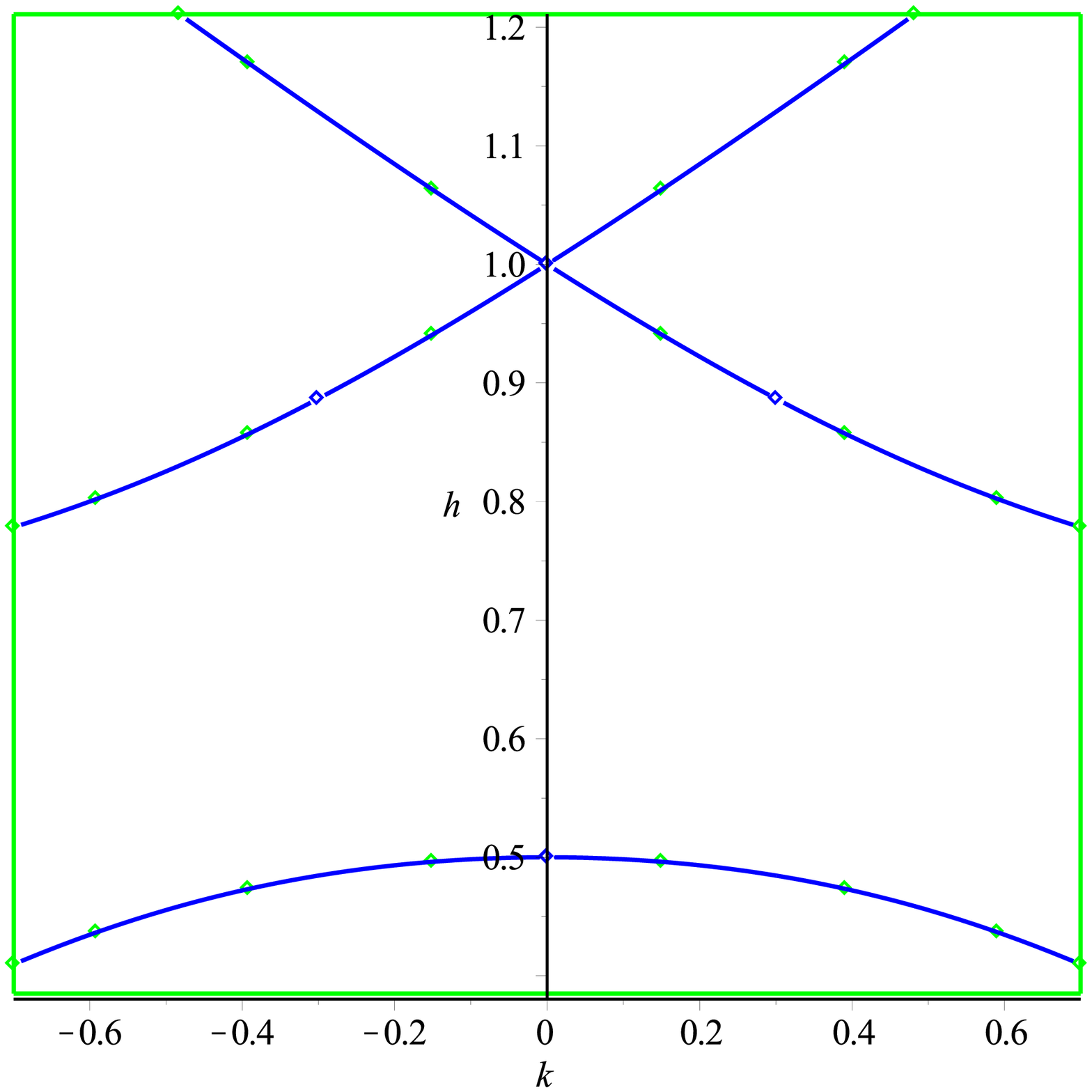}}
  \subfigure[]{\includegraphics[width=0.48\textwidth]{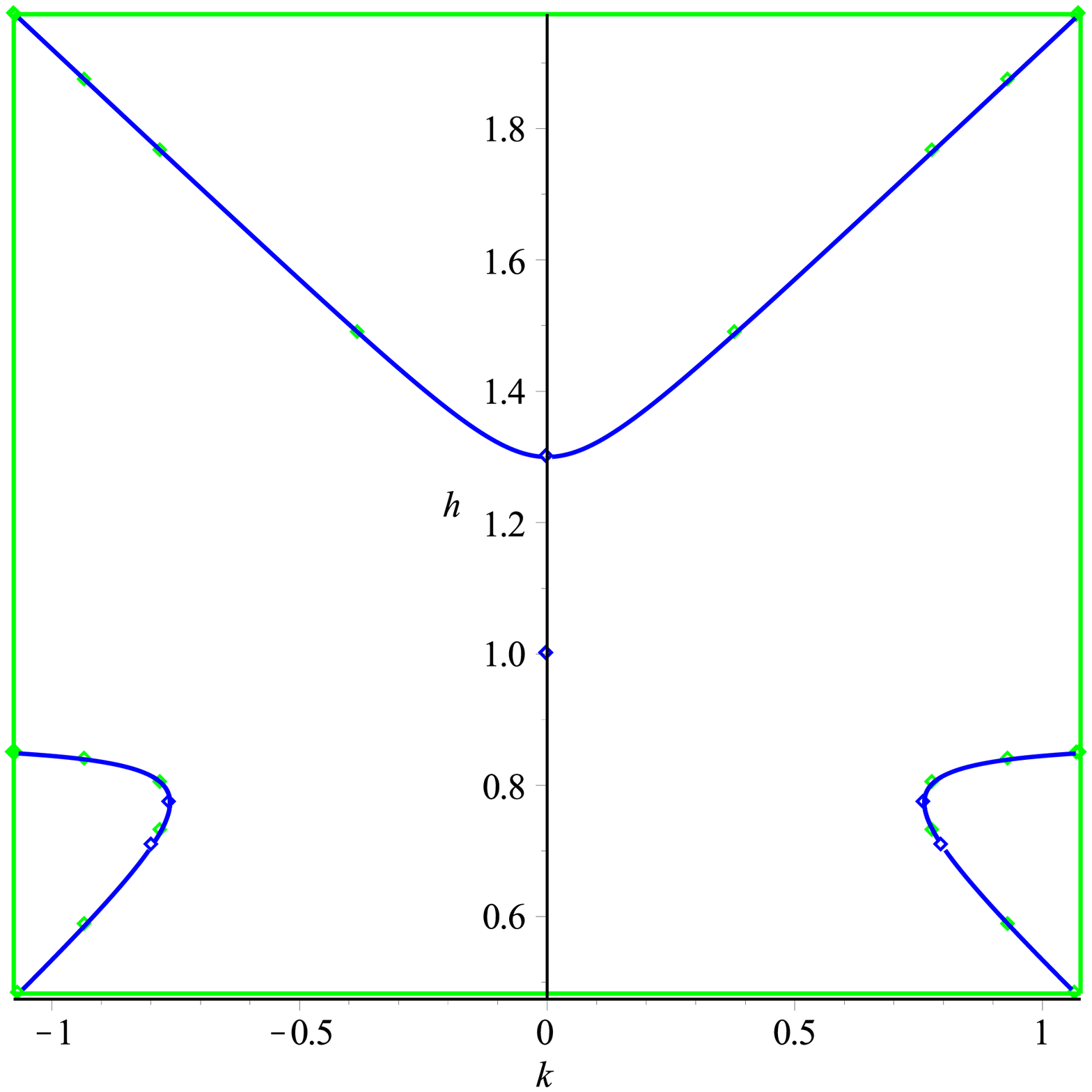}}
  \caption{\small\em Non-existence of solitary waves: (a) $\Fr\ =\ 0.5\ <\ 1$ and $\Bo\ =\ 0.25\ <\ \third\,$; (b) $\Fr\ =\ 1.3\ >\ 1$ and $\Bo\ =\ 0.5\ >\ \third\,$.}
  \label{fig:nothing}
\end{figure}

\begin{figure}
  \centering
  \subfigure[]{\includegraphics[width=0.48\textwidth]{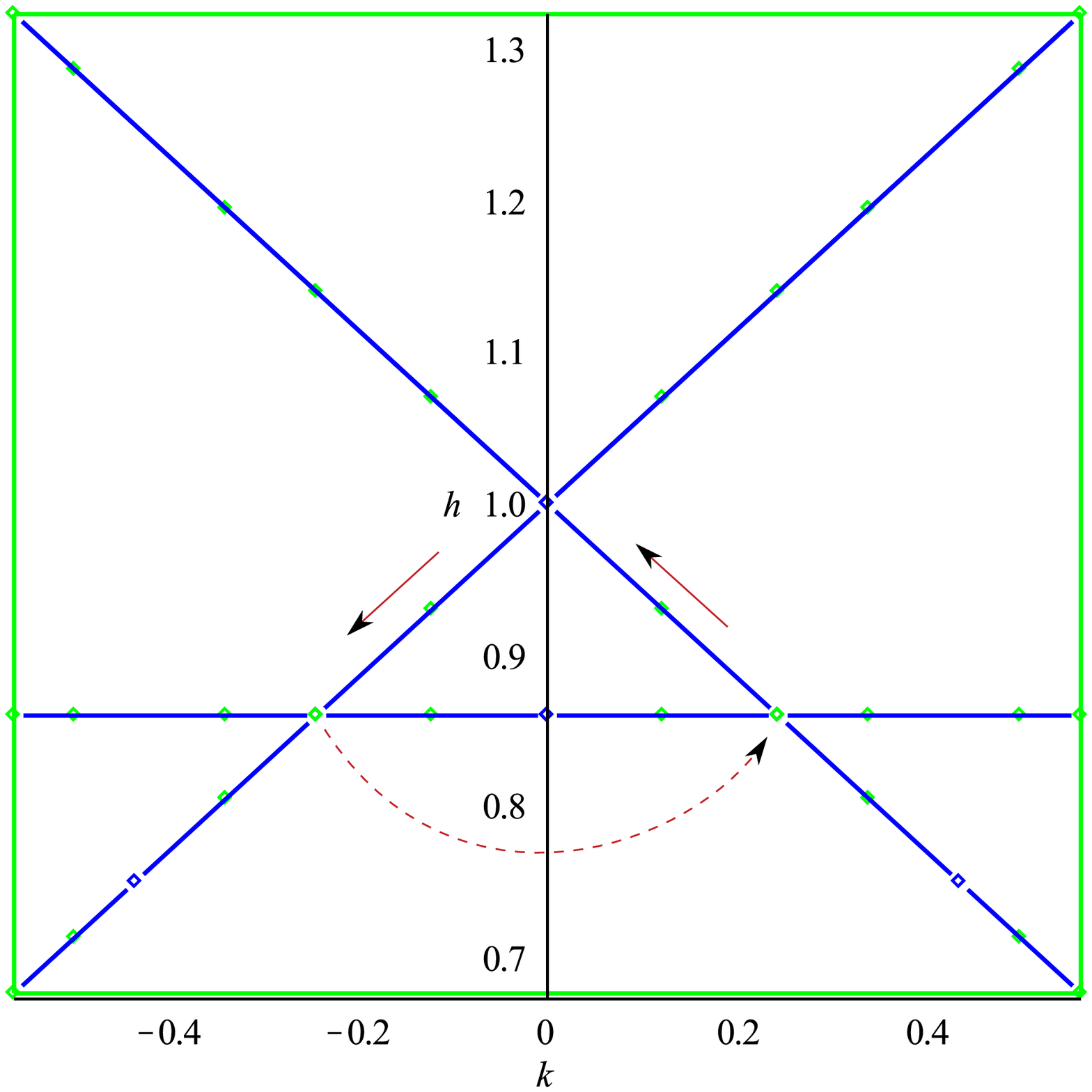}}
  \subfigure[]{\includegraphics[width=0.48\textwidth]{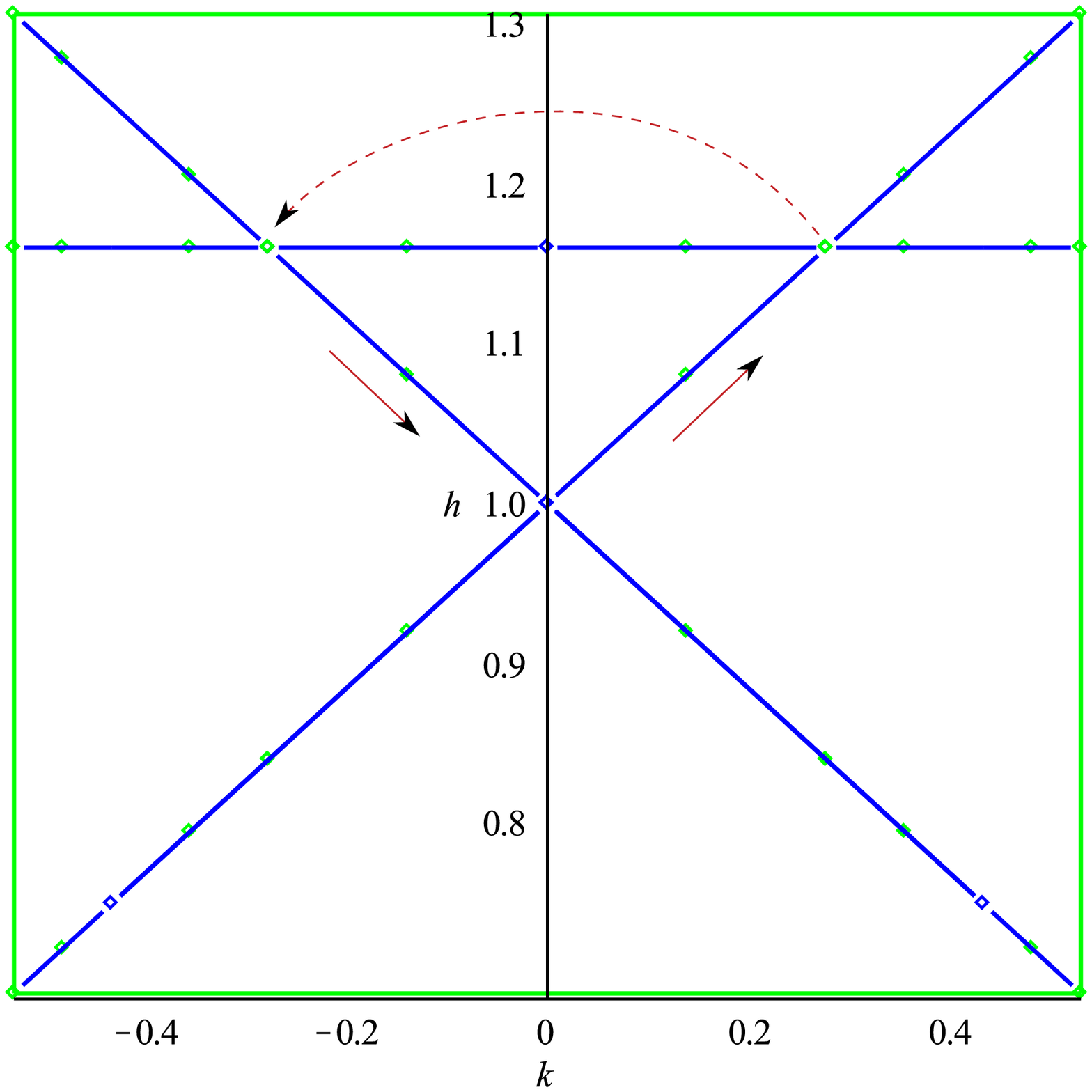}}
  \caption{\small\em Peakons in the critical case $\Bo\ =\ \third\,$: (a) $\Fr\ =\ 0.86\ <\ 1\,$; (b) $\Fr\ =\ 1.16\ >\ 1\,$. Arrows show the direction of motion along the branches and the dashed arrow represents a jump.}
  \label{fig:crit}
\end{figure}

Consider various characteristic values of parameters sampled from the configuration plane $(\Fr,\,\Bo)\,$. The case when $\Bo\ <\ \third$ and $\Fr\ >\ 1$ is depicted in Figure~\ref{fig:subcrit}. It corresponds to classical solitary waves of elevation and we call it a sub-critical case (since $\Bo$ is below the critical value). If $\Bo\ >\ \third$ and $\Fr\ <\ 1$ we have solitary waves of depression whose phase plane diagram is shown in Figure~\ref{fig:supercrit}. This case is referred to as the super-critical case. On the other hand, if $\Bo\ <\ \third$ and $\Fr\ <\ 1$ or $\Bo\ >\ \third$ and $\Fr\ >\ 1$ no solitary waves exist as it can be clearly seen on corresponding phase plane diagrams shown in Figure~\ref{fig:nothing}. However, the most intriguing situation is the critical case $\Bo\ =\ \third$. The corresponding phase plane diagrams are depicted in Figure~\ref{fig:crit}. These curves represent a new type of weakly singular solutions --- the so-called peakons \cite{Boyd1997, Liu2002}. The loop has to be discontinuous since there is a jump in the derivative at the wave crest. These jumps are indicated in Figure~\ref{fig:crit} with dashed arrows. The jump is possible since the derivative $\pd{\E}{h^{\,\prime}}\ =\ 0$ vanishes at the crest for $\Bo\ =\ \third$ and since there $\Fr\ =\ 1\ +\ a/d\,$. The phase plane of a peakon of depression is depicted in the left panel \ref{fig:crit}(\textit{a}) and a peakon of elevation in \ref{fig:crit}(\textit{b}). The peakons to SGN equations will be studied below using the means of the direct numerical simulation.

\begin{figure}
  \centering
 \includegraphics[width=0.99\textwidth, clip=true]{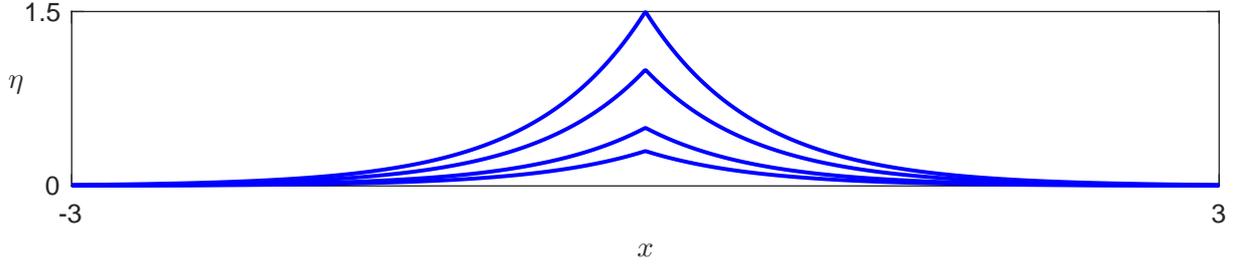}
  \caption{\small\em Peakons' profiles for Froude numbers $\Fr=1.3$, $1.5$, $2.0$ and $2.5$.}
  \label{fig:profiles}
\end{figure}

Figure~\ref{fig:profiles} shows the profiles of various peakons of elevation (since the analogous peakons of depression are the same but reflected about $\eta\ =\ 0$) for different phase speeds. It can be observed that the decay rate of the pulses is the same.

\section{Numerical study}

It is possible to derive analytically the free surface elevation corresponding to peakons discovered above using phase plane methods:
\begin{equation}\label{eq:anal}
  \eta_{\,p}\,(x,\,t)\ =\ a\,\ue^{-\sqrt{3}\,\abs{\xi}/d}\,, \qquad
  u_{\,p}\,(x,\,t)\ =\ \frac{c\,\eta_{\,p}\,(\xi)}{1\ +\ \eta_{\,p}\,(\xi)}\,, \qquad \xi\ =\ x\ -\ c\,t\,,
\end{equation}
where $c\ =\ \sqrt{1\ +\ a/d}$ is defined in \eqref{eq:speedampl} and in this derivation we used the results of the dispersion relation analysis for solitary waves \eqref{eq:disprel}. We notice that the wave amplitude $a\, >\, -d$ is a real free parameter and it is not necessarily positive. In agreement with the phase plane analysis, the peakons of depression are described by the same analytical expression with $-d\ <\ a\ <\ 0$ and the corresponding \textsc{Froude} number will be sub-critical, \ie $\Fr\ <\ 1\,$. We would like to underline that peakons are known to appear in various dispersionless limits \cite{Degasperis2002}.

In order to solve numerically the SGN equations \eqref{eq:serre1}, \eqref{eq:serre2} we use the standard \textsc{Galerkin} / Finite Element method with smooth cubic splines for the spatial discretization (of fourth order of accuracy) combined with the classical four stage, fourth order, explicit \textsc{Runge}--\textsc{Kutta} method for the time stepping \cite{Mitsotakis2014, Antonopoulos2017}. The boundary conditions are periodic in the simulations below. However, we took the domains sufficiently large in order to avoid  interactions with the boundaries when possible. We mention also that the method degenerates to the first order of accuracy at the singular crest.

\begin{figure}
  \centering
  \includegraphics[width=0.99\textwidth, clip=true]{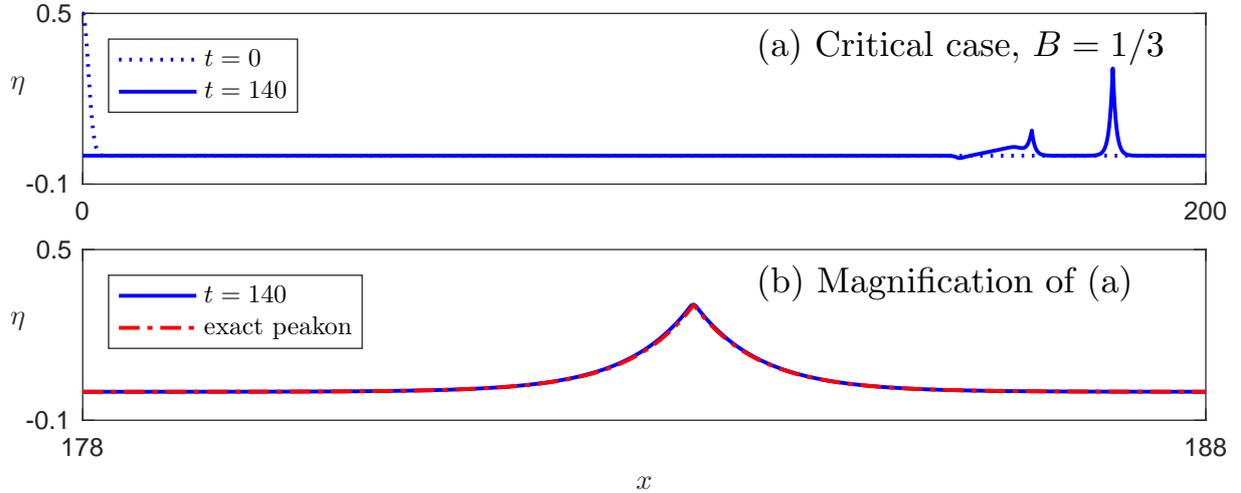}
  \caption{\small\em Emergence of a peakon from a smooth initial condition.}
  \label{fig:peakon}
\end{figure}

\begin{figure}
  \centering
  \includegraphics[width=0.99\textwidth, clip=true]{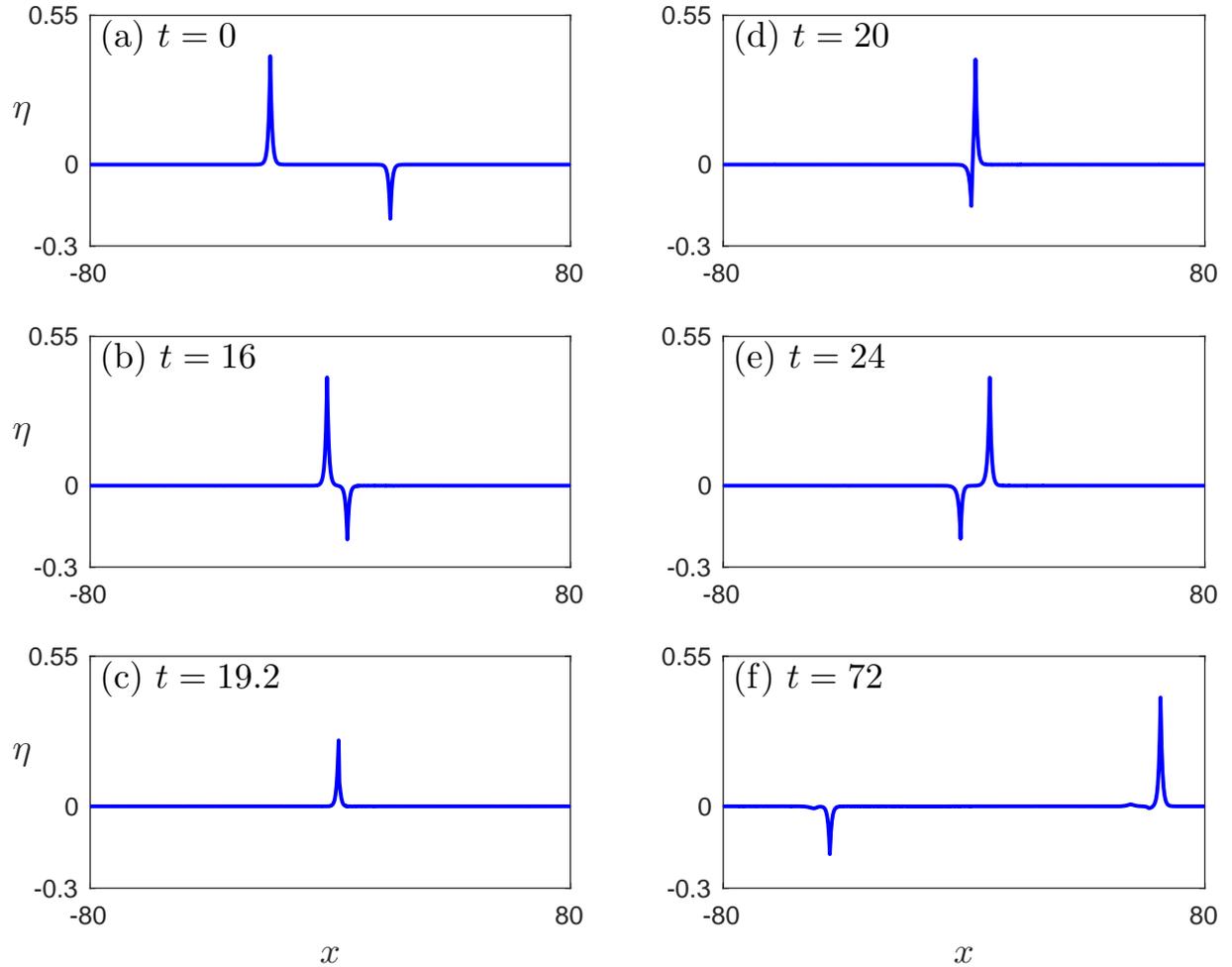}
  \caption{\small\em A head-on collision of two peakons of elevation ($a/d = 0.4$) and depression ($a/d = -0.2$).}
  \label{fig:headon}
\end{figure}

\begin{figure}
  \centering
  \includegraphics[width=0.99\textwidth, clip=true]{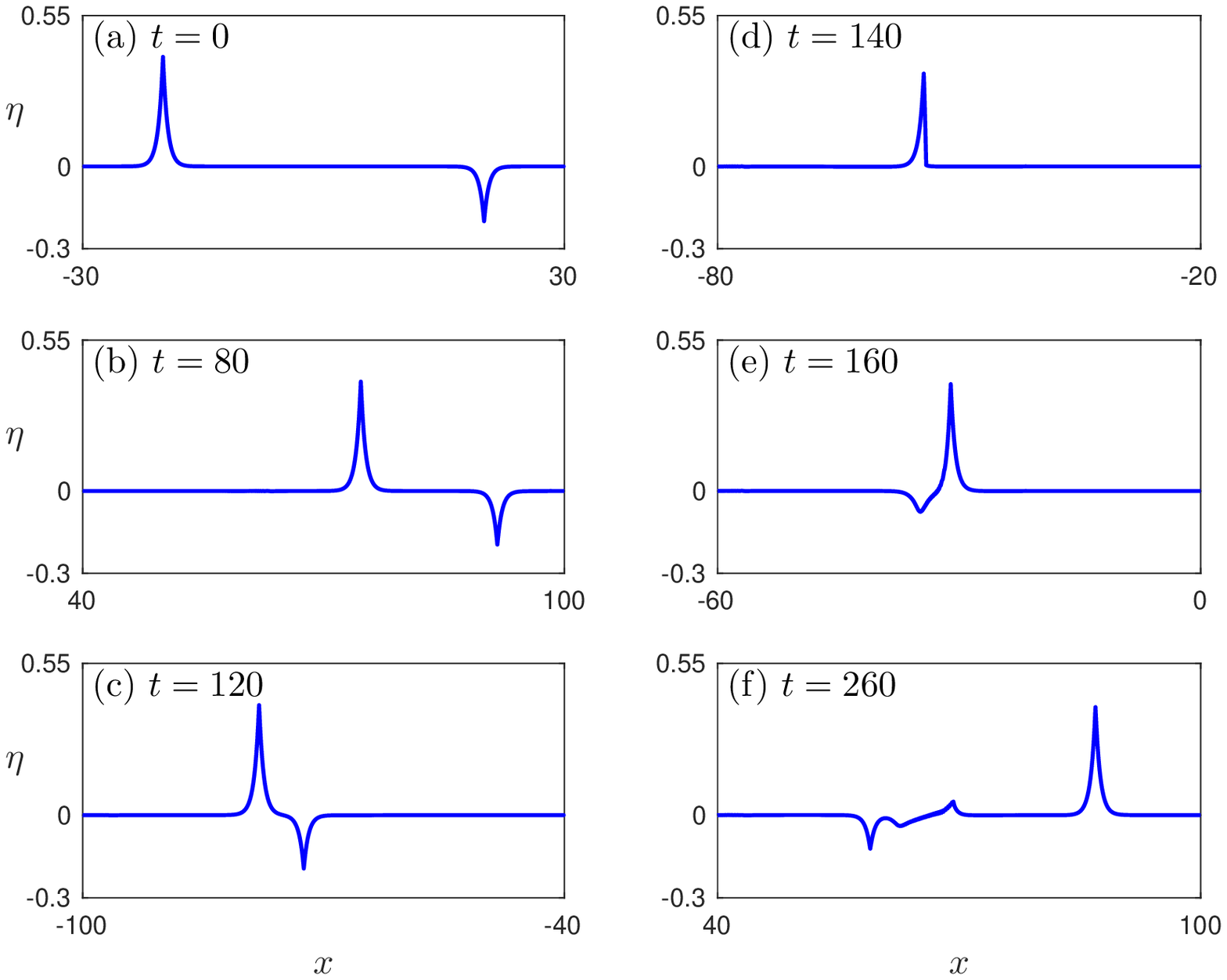}
  \caption{\small\em An overtaking collision of two peakons of elevation ($a/d = 0.4$) and depression ($a/d = -0.2$).}
  \label{fig:over}
\end{figure}

The first numerical experiment will consist in starting with a smooth initial condition in the critical case $\Bo\ =\ 1/3$
\begin{equation*}
  h(x,\,0) = 1 + \frac{1}{2}\,\ue^{-\half\;x^2}\,, \quad
  u(x,\,0) = 0\,, \quad x \in [-200,\,200]\,.
\end{equation*}
The evolution of this initial condition is shown in Figure~\ref{fig:peakon}. One can see that a peakon emerges under the SGN dynamics in the critical regime. In order to support this claim, the lower panel of Figure~\ref{fig:peakon} shows a magnification of the free surface with a peakon given by the analytical formula \eqref{eq:anal} superimposed on the numerical peakon. The amplitude of the emerged peakon is $a/d\ \approx\ 0.3058\,$. The spatial mesh used in this computation is $\Delta x\ =\ 0.02$ and the time step was taken to be $\Delta t\ =\ 0.002\,$. It is noted that these numerically generated waves from the evolution of a smooth initial condition belong into the finite element space and they propagate without shedding any spurious oscillations and they remain smooth approximations of the exact peakons. If we start with a different initial condition, the generated peakon will have a different amplitude. For instance, if the initial displaced mass is negative (\ie we start with a trough rather than a bump), we shall have a peakon of depression. Moreover, if the computations are continued for sufficiently long time and the initial displaced mass is sufficiently large, we can see the emergence of a finite number of peakons. Thus, in some sense the dynamics of SGN equations are similar to the \textsc{Korteweg}--\textsc{de Vries} (KdV) equation where smooth solitary waves (of elevation) emerge \cite{Segur2} and also similar to the \textsc{Camassa}--\textsc{Holm} (CH) equation where peakons emerge, \cite{Camassa1993}. However, there  are important differences in the behaviour of the SGN, KdV and CH equations. The former in the critical case supports travelling waves of elevation \emph{and} depression at the same time. The latter models are integrable. It is not the case of the SGN equations. For this reason we performed additional numerical experiments of head-on (see Figure~\ref{fig:headon}) and overtaking (see Figure~\ref{fig:over}) collisions. We considered peakons of different polarities and very fine meshes ($\Delta x\ =\ 0.004\,$, $\Delta t\ =\ 0.0004$) to simulate this process very accurately. In particular, one can see that these collisions are inelastic, which indicates the non-integrability of SGN equations \cite{Ablowitz2011}. It is noted that the head-on collision results to a new peakon of depression propagating with larger phase speed than the initial one and a new peakon of elevation propagating with smaller phase speed. The small amplitude tails generated by the peakon of depression propagate faster than the peakon itself contrary to the small amplitude tails generated by the peakon of elevation. The overtaking collision is more effective to the small amplitude peakon, which is actually resolved into several waves including peakons of elevation and depression. The large amplitude peakon that emerges after the interaction is very similar to the initial one and remained stable. It is worth to mention that the trailing tails resulting from the interactions of peakons can in principle evolve into secondary peakons have the shape of a $N-$shaped wavelets rather than the usual form of a trailing dispersive tail. These wavelets could in principle evolve into secondary peakons, if we allowed the simulations to continue for larger times. Similar results were observed when we considered overtaking interactions between peakons of elevation. The overtaking collision though resulted in peakons of elevation of very similar amplitude with the initial pulses in addition to small amplitude trailing tails.

\begin{figure}
  \centering
  \bigskip
  \includegraphics[width=0.89\textwidth]{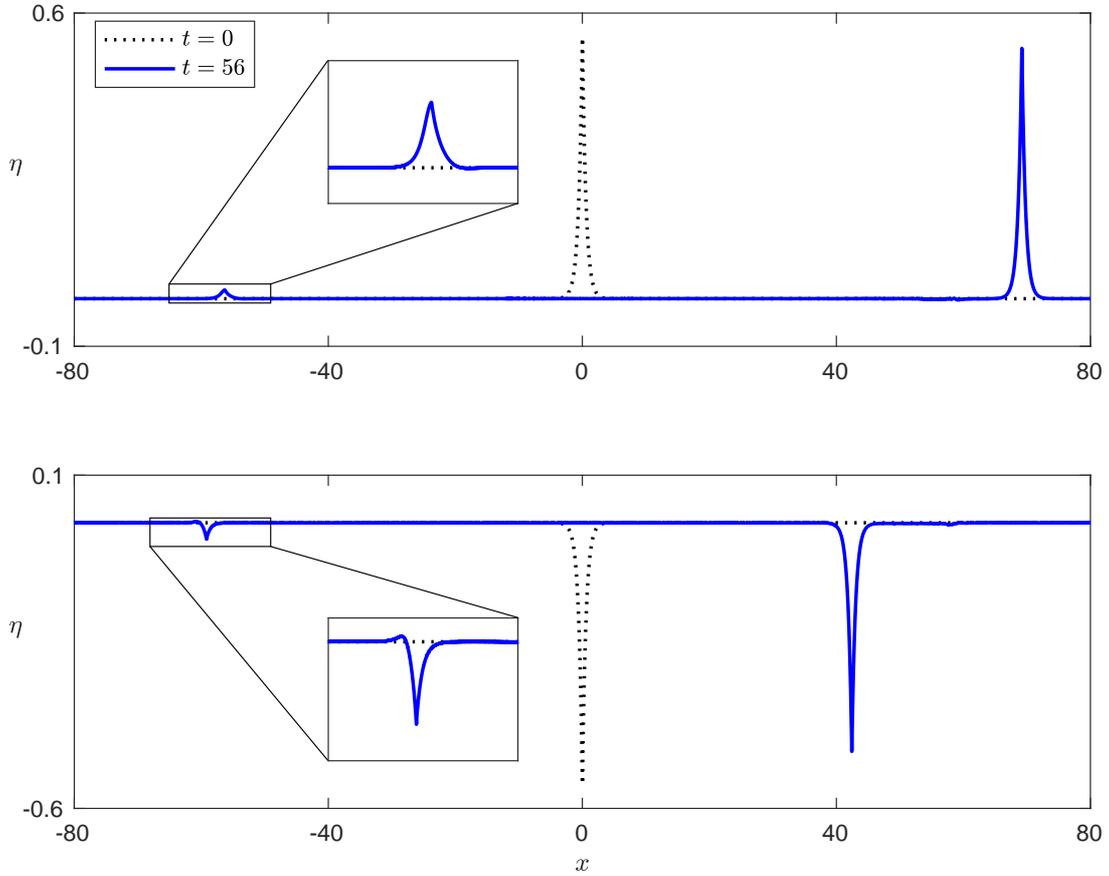}
  \caption{\small\em Evolution of a perturbed peakon under SGN equations dynamics: (a) perturbed peakon of elevation with initial amplitude $a\ =\ 0.5\,$; (b) perturbed peakon of depression with $a\ =\ -0.5\,$. The initial condition is shown with black dotted lines and the final state with blue solid lines.}
  \label{fig:stab}
\end{figure}

In order to shed some light onto peakon stability properties, we considered also another numerical experiment inspired by a previous study \cite{DDMM}, where we take an exact peakon $\bigl(\eta_{\,p}\,(x,\,0),\,u_{\,p}\,(x,\,0)\bigr)$ of amplitude $a$ (in our simulations we take $a\ =\ \pm\,\half$) given by formula \eqref{eq:anal}. Then, the free surface elevation is perturbed by the amount of $10\%\,$, while the initial velocity is kept the same:
\begin{equation*}
  h_{\,0}\,(x)\ =\ 1\ +\ 1.1\times\eta_{\,p}\,(x,\,0)\,, \qquad
  u_{\,0}\,(x)\ =\ u_{\,p}\,(x,\,0)\,.
\end{equation*}
The initial conditions for the free surface elevation are depicted in Figure~\ref{fig:stab} with black dotted lines. We computed the evolution of these initial conditions under the SGN dynamics up to $t\ =\ 56\,$. The final states of the free surface elevation are shown in Figure~\ref{fig:stab} with the blue solid lines. One can see the initial peakons with slightly perturbed amplitudes ($\tilde{a}_{\,1}^{\,-}\ \approx\ -\,0.4804$ and $\tilde{a}_{\,1}^{\,+}\ \approx\ 0.5258$ instead of $a\ =\ \pm\,0.5$ correspondingly). However, one can notice also secondary peakons of amplitudes $\tilde{a}^{\,+}_{\,2}\ \approx\ 1.23\times 10^{-2}$ and $\tilde{a}^{\,-}_{\,2}\ \approx\ -\,3.47\times 10^{-2}\,$. If we allow more time for the propagation, other coherent structures of smaller amplitude might appear. This computation shows that perturbed peakons inevitably evolve into nearby peakons as well, indicating that peakons to SGN equations are orbitally attractive.

It is noted that we considered the $L^2$ projection of the exact peakons onto the finite element space as numerical initial conditions. Numerically approximated initial peakon shed a negligible artifact which is analogous to the accuracy of the numerical method and does not affect the numerical results. Since these artifacts do not propagate they can be easily cleaned from the rest of the solution using the cleaning technique described in \cite{DDMM}.


\section{Conclusions}

The physical pertinence of the proposed solutions has to be further investigated by studying the full \textsc{Euler} equations in the critical regime as well. The last problem is more complicated, since the variety of known travelling capillary--gravity waves is huge \cite{Vanden-Broeck2010, Clamond2015a}. Moreover, even if the employed model can be derived in a physical manner using elegant variational techniques, its pertinence is restricted due to the small-slope assumption. However, we presented enough evidence that the mathematical model under consideration admits a continuous family of peaked solitary waves both of elevation \emph{and} depression. It is shown analytically that the decay rate of these peakons is universal and does not depend on the solution amplitude. Additionally, these solutions follow the traditional speed--amplitude relation \eqref{eq:speedampl}, which coincides with the empirical relation of Scott~\textsc{Russel} for smooth solitary waves. The dynamics of these waves were studied. These results apply also to \emph{line solitary waves} in 3D. Finally, we showed that these solutions are stable in the sense of dynamical systems, \ie peakons emerge inevitably from `arbitrary' smooth initial conditions after a short transient period of time. The amplitude of the emerged peakon seems to depend on the mass of the initial condition with respect to the still water level. As the main conclusion of this Letter we state that peakons are \emph{emerging} coherent structures in the critical regime of the SGN equations \eqref{eq:serre1} -- \eqref{eq:serre2}.


\subsection*{Acknowledgments}
\addcontentsline{toc}{subsection}{Acknowledgments}

The work of D.~\textsc{Mitsotakis} was supported by the Marsden Fund administrated by the Royal Society of New Zealand. A.~\textsc{Assylbekuly} and D.~\textsc{Zhakebayev} acknowledge the support from Science Committee of MES of the Republic of Kazakhstan under the project N$^\circ$ 2018/GF4.


\addcontentsline{toc}{section}{References}
\bibliographystyle{abbrv}
\bibliography{biblio}

\end{document}